\documentclass[a4paper,aps,prd,onecolumn,preprintnumbers,showpacs,nofootinbib]{revtex4}
\usepackage{amsmath,amssymb,graphics,epsfig,subfigure}
\usepackage{color}

\usepackage[
      colorlinks=true,
      linkcolor=blue,
      urlcolor=blue,
      filecolor=black,
      citecolor=blue,
            ]{hyperref}

\def\be{\begin{equation}}
\def\ee{\end{equation}}
\def\beq{\begin{eqnarray}}
\def\eeq{\end{eqnarray}}

\newcommand{\m}{\mathrm}

\renewcommand{\[}{\left[}
\renewcommand{\]}{\right]}

\begin{document}
\renewcommand{\baselinestretch}{1.3}

\title{Critical phenomena in the extended phase space of Kerr-Newman-AdS black holes}

\author{Peng Cheng \footnote{pcheng14@lzu.edu.cn},
        Shao-Wen Wei \footnote{weishw@lzu.edu.cn,  corresponding author},
        Yu-Xiao Liu \footnote{liuyx@lzu.edu.cn, corresponding author}}

\affiliation{Institute of Theoretical Physics, Lanzhou University, Lanzhou 730000, People's Republic of China}

\begin{abstract}
Treating the cosmological constant as a thermodynamic pressure, we investigate the critical behavior of a Kerr-Newman-AdS black hole system. The critical points for the van der Waals like phase transition are numerically solved. The highly accurate fitting formula for them is given and is found to be dependent of the charge $Q$ and angular momentum $J$. In the reduced parameter space, we find that the temperature, Gibbs free energy, and coexistence curve depend only on the dimensionless angular momentum-charge ratio $\epsilon=J/Q^2$ rather than $Q$ and $J$. Moreover, when varying $\epsilon$ from 0 to $\infty$, the coexistence curve will continuously change from that of the Reissner-Nordstr\"{o}m-AdS black hole to the Kerr-AdS black hole. These results may guide us to study the critical phenomena for other thermodynamic systems with two characteristic parameters.
\end{abstract}



\maketitle

\section{Introduction}

Thermodynamic of black holes has been a stirring topic in gravitational physics since the seminal start in 1970's \cite{Bardeen1973, Bekenstein:1973ur}. After that, the stable large black hole to thermal gas phase transition was discovered by Hawking and Page \cite{Hawking1983}. The thermodynamic properties and phase transitions for the asymptotically anti-de Sitter (AdS) black holes have attracted a wide attention \cite{Chamblin:1999hg, Chamblin:1999tk, Cvetic:1999ne}, in light of the AdS/CFT correspondence \cite{Maldacena:1997re, Gubser:1998bc, Witten:1998qj} which admits a gauge duality description via a dual thermal field theory. The in-depth studies for the charged AdS black holes show that Hawking-Page phase transitions exist in the grand canonical ensemble with fixed electric potential. And there are small-large black holes phase transitions, which are analogue to the liquid-gas transitions of the van der Waals (vdW) fluid, existing in the canonical ensemble with fixed charge \cite{Banerjee2011, Banerjee:2011raa, Niu:2011tb}. The similar behaviors were then found for the Kerr-AdS and Kerr-Newman (KN)-AdS black holes \cite{Caldarelli:1999xj, Tsai:2011gv}, as well as other AdS black holes.

Recently, the precise analogy of an AdS black hole system and a vdW fluid was completely established by regarding the variation of the cosmological constant as a thermodynamic pressure and its conjugate quantity as a volume \cite{Kastor:2009wy, Dolan:2010ha, Dolan:2011xt, Kubiznak:2012wp}. There are many new features in the extended phase space, for example, the equation of state can be used for comparison with ordinary thermodynamic systems \cite{Kubiznak:2012wp}. Moreover, the thermodynamic volume was shown to satisfy the reverse isoperimetric inequality \cite{Cvetic:2010jb}, and one could obtain a more fundamental theory that admits the variation of the cosmological constant \cite{Kastor:2009wy}. Many works demonstrated that there extensively exists a small-large black hole phase transition in the AdS black hole systems \cite{Hendi:2012um, Chen:2013ce, Zhao:2013oza, Cai:2013qga, Spallucci:2013osa, Spallucci:2013jja, Xu:2013zea, Zou:2013owa, Mo:2014qsa, Mo:2014mba, Zou:2014mha, Mirza:2014xxa, Dolan:2014vba, Xu:2014tja, Xu:2015rfa, Dolan:2014jva, Zhang:2015ova, Wei:2014qwa, Wei:2015ana}. Analogy to a vdW fluid system, they not only share the same oscillatory behavior of the isothermal line, and the swallow tail behavior of the Gibbs free energy, but also share the same critical exponents and scaling laws near the critical point. Moreover, there also exhibit some interesting multi-critical phenomena, such as the reentrant phase transition and triple point \cite{Gunasekaran:2012dq, Wei:2014hba, Frassino:2014pha, Sherkatghanad:2014hda, Altamirano:2013ane, Altamirano:2013uqa, Altamirano:2014tva}.

A further analogy between the vdW fluid and AdS black hole was made in Ref. \cite{Wei:2015iwa}, where a new concept, the number density of black hole molecules, was introduced. The difference of the number density between the small and large black hole phases naturally provides us with an order parameter to describe the phase transition. When an AdS black hole system crosses the coexistence curve, there is a nonvanishing latent heat, just like an ordinary thermodynamic system. It was also revealed that there exists interaction between two black hole molecules. These provides a new perspective for the phase transition of AdS black holes.

Among these investigations, one of the interesting phenomena is that, in the reduced parameter space, the critical behaviors of the Reissner-Nordstr\"{o}m (RN)-AdS and Kerr-AdS black holes are independent of the charge $Q$ and angular momentum $J$, respectively. The reason, according to the suggestion of Ref. \cite{Wei:2015ana}, is that they are single characteristic parameter thermodynamic systems. Here we would like to ask, for a black hole with nonvanishing charge $Q$ and angular momentum $J$, how does this case behave? According to our classification of the thermodynamic quantities \cite{Wei:2015ana}, a KN-AdS black hole is a two-characteristic-parameter thermodynamic system. Therefore, its critical behavior should be related both to $Q$ and $J$. After a numerical calculation, the critical points indeed depend on $Q$ and $J$. However, in the reduced parameter space, the temperature and Gibbs free energy only depend on a dimensionless angular momentum-charge (AMC) parameter $\epsilon$ rather than $Q$ and $J$. Moreover, the coexistence curve is also dependent of $\epsilon$ only. And the coexistence curve continuously changes from that of the RN-AdS black hole to the Kerr-AdS black hole, as varying $\epsilon$ from 0 to $\infty$.

The main purpose of this paper is to study the critical phenomena for a KN-AdS black hole in the extend phase space. The paper is organized as follows. We review the thermodynamic properties of 4-dimensional RN-AdS and Kerr-AdS black holes, especially their thermodynamic properties in the reduced parameter space in Sec. \ref{RN-Kerr}. In Sec. \ref{Kerr-Newman}, we study the thermodynamic critical behaviors of KN-AdS black holes. The equation of state, critical quantities, and the $\tilde T-\tilde S$, $\tilde G-\tilde T$ diagrams are obtained. We find that all the isothermal and Gibbs free energy diagrams depend only on $\epsilon$ rather than on $Q$ and $J$, so does the coexistence curve. Finally, we summarize and discuss our results in Sec. \ref{conclusions}. We use geometric units of $c=G_d=\hbar =k_B=1$ throughout the paper.

\section{Critical behaviors of RN-AdS black hole and Kerr-AdS black hole}\label{RN-Kerr}
Before investigating the KN-AdS black hole system, we would like to give a short review of the critical behaviors for the RN-AdS black hole and Kerr-AdS black hole.

\subsection{RN-AdS black holes}
\begin{figure}[htb]
    \subfigure[]{\label{RNTS}
    \includegraphics[height=0.3\textwidth]{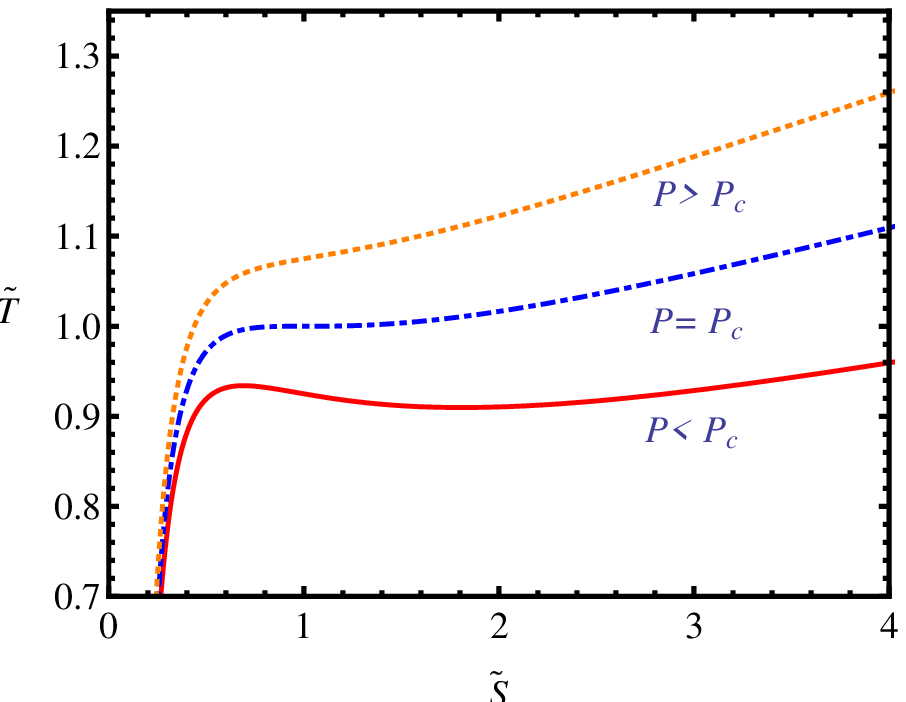}}
    \subfigure[]{\label{RNGT}
    \includegraphics[height=0.3\textwidth]{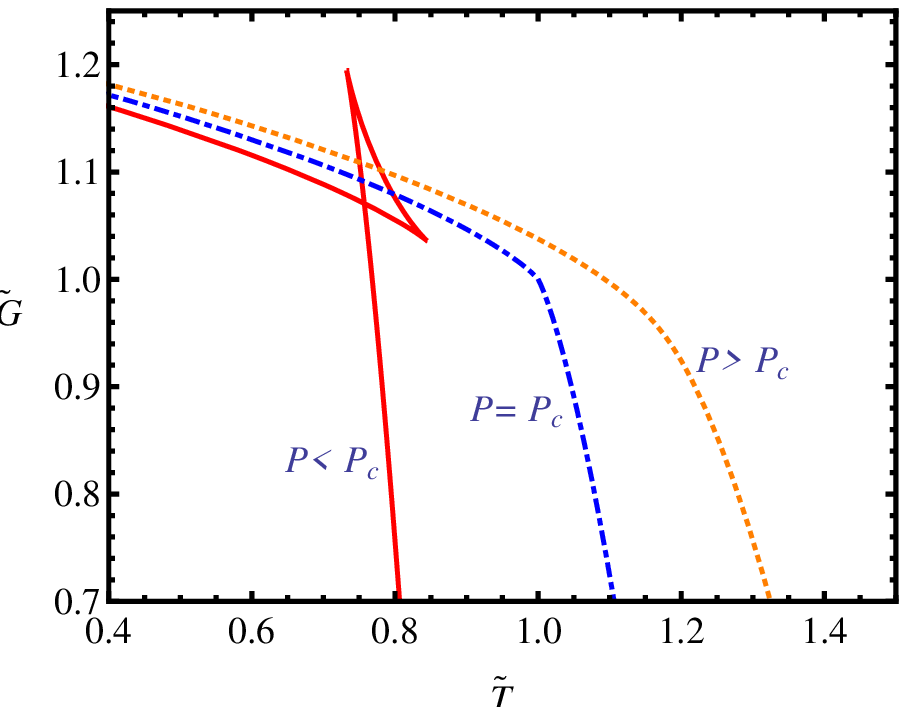}}
    \caption{Isobaric lines for the RN-AdS black hole. The dotted, dot-dashed, and solid lines correspond to $\tilde{P}$= 1.5, 1, and 0.5, respectively. (a) $\tilde{T}-\tilde{S}$ diagram. (b) $\tilde{G}-\tilde{T}$ diagram.}\label{RNPvGT}
\end{figure}

The solution of a 4-dimensional spherical RN-AdS black hole can be written as
\beq
ds^2 &=& -f(r) \m{d}t^2 + \frac{\m{d}r^2}{f(r)} + r^{2} \m{d}\Omega_{2}^2\,,\nonumber\\
F&=&\m{d}A\,,\quad A=-\frac{q}{r}\m{d}t\,,
\eeq
where $\m{d}\Omega_2^2$ stands for the standard element on $S^2$ and the $f(r)$
is given by
\be\label{RNmet}
f(r)= 1 - \frac{M}{r} + \frac{Q^2}{r^{2}} + \frac{r^2}{l^2}\,.
\ee
The parameters $M$ and $Q$ are the mass and charge of the black hole. Following the idea of \cite{Dolan:2010ha}, the cosmological constant $\Lambda$ can be treated as a thermodynamic pressure $P$,
\be
P=-\frac{\Lambda}{8\pi}=\frac{3}{8 \pi l^2}.
\ee
Making an analogy with the vdW fluid, there exists a specific volume $v$ for the black hole \cite{Kubiznak:2012wp}
\be
v=2l_P^2r_+\,,
\ee
where $l_P$ is the Planck length and equals to 1 in geometric units. The temperature and Gibbs free energy for the black hole can be written as \cite{Kubiznak:2012wp}
\beq
T &=& \frac{8 P S^2-\pi  Q^2+S}{4 \sqrt{\pi } S^{3/2}}\,,\label{RNTe}\\
G &=& \frac{S (3-8 P S)+9 \pi  Q^2}{12 \sqrt{\pi } \sqrt{S}}\,.\label{RNGi}
\eeq
The critical point can be obtained through solving $\partial_{v}P=\partial_{v,v}P=0$, or $\partial_{S}T=\partial_{S,S}T=0$,
which lead to
\be\label{RNcritical}
P_c=\frac{1}{96\pi Q^2}\,,\quad S_c=6 \pi  Q^2\,,\quad T_c=\frac{\sqrt{6}}{18\pi Q}\,,\quad v_c=2\sqrt{6} Q\,,\quad G_c=\frac{\sqrt 6}{3}Q\,.
\ee
Here, we define the thermodynamic quantities in the reduced parameter space as
\beq
\tilde{T} =\frac{T}{T_c},~~~\tilde{P} =\frac{P}{P_c},~~~\tilde{S} =\frac{S}{S_c},~~~\tilde{G} =\frac{G}{G_c}\,.
\eeq
Therefore, the temperature and Gibbs free energy in the reduced parameter space can be expressed as \cite{Wei:2014qwa}
\beq
\tilde{T} &=& \frac{3\tilde{P} \tilde{S}^2+6\tilde{S}-1}{8 \tilde{S}^{3/2}}\,,\\
\tilde{G} &=& \frac{-\tilde{P} \tilde{S}^2+6 \tilde{S}+3}{8 \sqrt{\tilde{S}}}\,.
\eeq
It is easy to find that these quantities are charge $Q$ independent, and are described in Fig.\ref{RNPvGT} for the different values of the pressure $P$. For $P<P_{c}$, there exist an oscillatory behavior of $\tilde T$ and a characteristic swallow tail behavior of $\tilde G$. Both of them imply a firs-order phase transition for the black hole system. While for $P>P_{c}$, the oscillatory behavior and characteristic swallow tail disappear, which reveals no phase transition occurring.

\begin{figure}[htb]
    \subfigure[]{\label{RNPT}
    \includegraphics[height=0.4\textwidth]{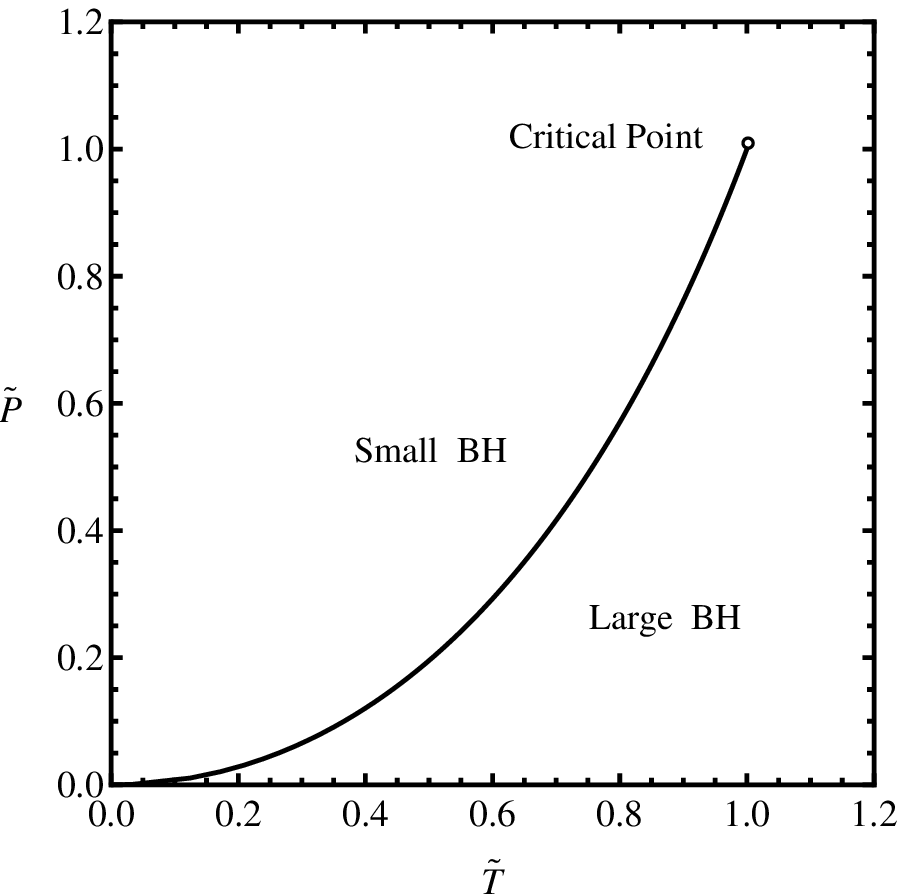}}
    \subfigure[]{\label{KerrPT}
    \includegraphics[height=0.4\textwidth]{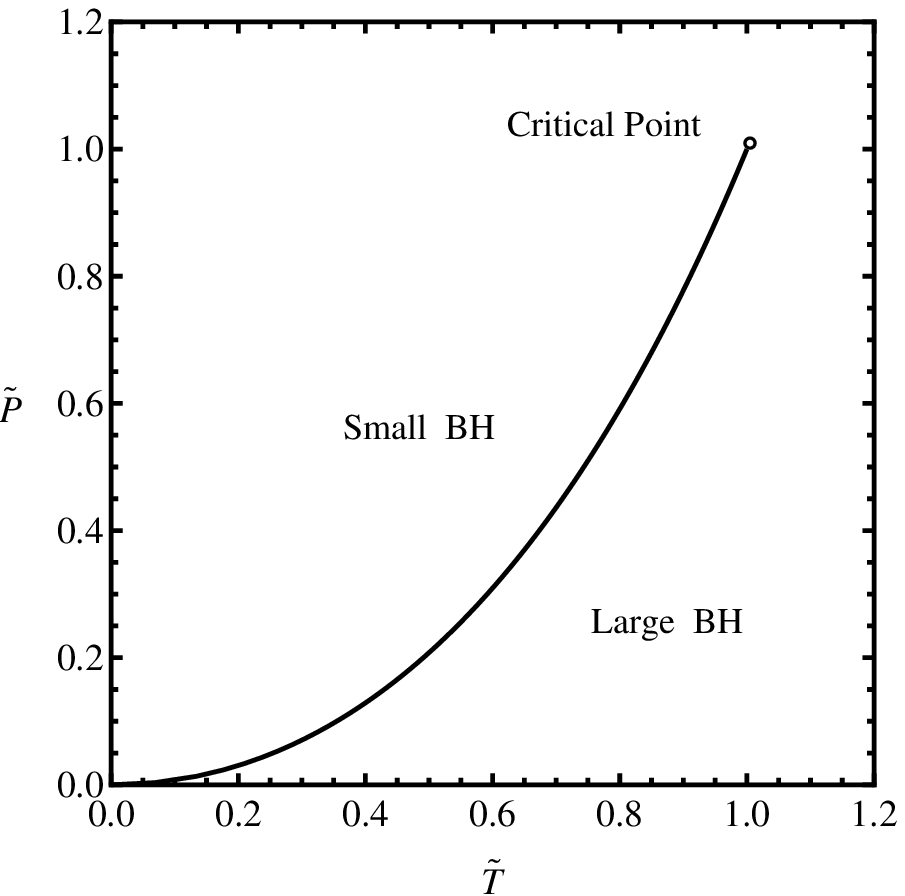}}
    \caption{Coexistence curves in the reduced parameter space for (a) the RN-AdS black hole and (b) the Kerr-AdS black hole. Both the curves are independent of the charge $Q$ and angular momentum $J$, respectively.}
\end{figure}

Since the reduced temperature and Gibbs free energy are both free of the charge $Q$, the reduced coexistence curve is also free of $Q$. For the $4$-dimensional RN-AdS black hole system, the authors of Ref. \cite{Lan:2015bia} firstly showed that there exists an analytical form for the coexistence curve
\be\label{ma}
 \tilde P=\[1-2\cos\left(\frac{\arccos(1-\tilde T^2)+\pi}{3}\right)\]^2,\quad \tilde T\in(0,\;1)\,.
\ee
We clearly show the curve in Fig. \ref{RNPT}. For higher dimensional RN-AdS black holes, only numerical results exist. And the fitting formula of the coexistence curves can be found in \cite{Wei:2014qwa}.

\subsection{Kerr-AdS black holes}

\begin{figure}[htb]
    \subfigure[]{\label{KerrTS}
    \includegraphics[height=0.3\textwidth]{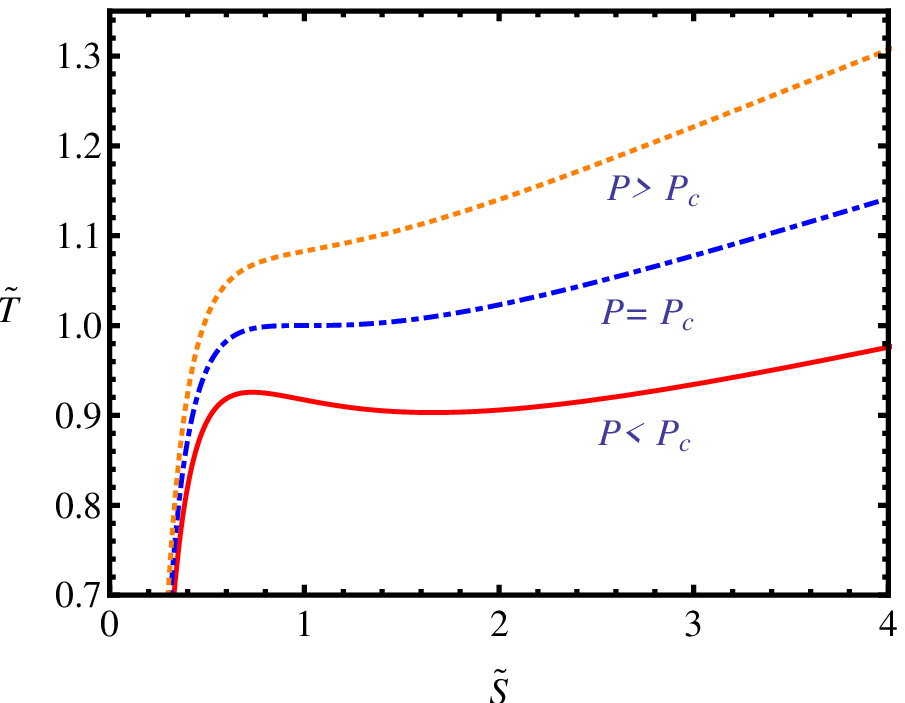}}
    \subfigure[]{\label{KerrGT}
    \includegraphics[height=0.3\textwidth]{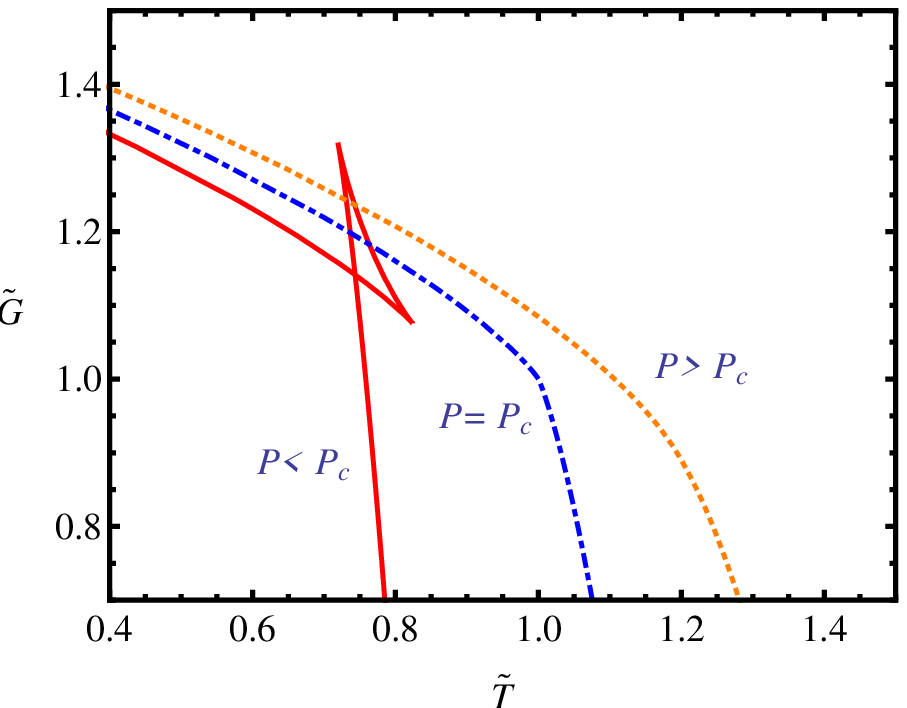}}
    \caption{Isobaric lines for the Kerr-AdS black hole. The dotted, dot-dashed, and solid lines correspond to $\tilde{P}$= 1.5, 1, and 0.5, respectively. (a) $\tilde{T}-\tilde{S}$ diagram. (b) $\tilde{G}-\tilde{T}$ diagram.} \label{KerrTSGT}
\end{figure}

The 4-dimensional line element of a Kerr-AdS black hole in Boyer-Lindquist-like coordinates $(t, r, \theta, \phi)$ is given by
\beq\label{Kerrmetric}
ds^2 &=& - {\Delta_r \over \rho^2}\Bigr(\,\m{d}t \; - \; {a \m{sin} \over \Xi}^2\theta \,\m{d}\phi\Bigl)^2\;+\;{\rho^2 \over \Delta_r}\m{d}r^2\;+\;{\rho^2 \over \Delta_{\theta}}\m{d}\theta^2  \nonumber \\
&~&+\;{\m{sin}^2\theta \,\Delta_{\theta} \over \rho^2}\Bigr(a\,\m{d}t \; - \;{r^2\,+\,a^2 \over \Xi}\,\m{d}\phi \Bigl)^2,
\eeq
where
\beq
\rho^2& = & r^2\;+\;a^2\m{\cos}^2\theta, \nonumber\\
\Delta_r & = & (r^2+a^2)\Big(1 + {r^2\over l^2}\Big) - 2mr,\nonumber\\
\Delta_{\theta}& = & 1 - {a^2\over l^2} \, \m{\cos}^2\theta, \nonumber\\
\Xi & = & 1 - {a^2\over l^2}.
\eeq
Regarding the cosmological constant as a pressure $P$, the temperature $T$ and Gibbs free energy $G$ can be expressed in terms of the entropy $S$, angular momentum $J$, and pressure $P$:
\beq
T &=& \frac{S^2 (64P^2 S^2 +32PS +3) -12\pi^2 J^2}{4\sqrt{\pi}S^{3/2}\sqrt{8P S +3} \sqrt{12\pi^2 J^2+S^2(8P S +3)}},\\
G &=& \frac{12\pi^2J^2 (16 PS +9) -64P^2 S^4 +9S^2}{12\sqrt{\pi} \sqrt{S} \sqrt{8PS +3} \sqrt{12\pi^2J^2 +S^2(8PS +3)}}\,.
\eeq
In Ref. \cite{Wei:2015ana}, we fund the solution of the analytical critical point for the Kerr AdS black hole by dimensional analysis:
\beq
P_c = \alpha \cdot J^{-1},~~
v_c = \beta \cdot J^{1/2},~~
T_c = \gamma \cdot J^{-1/2},~~
G_c =\delta \cdot J^{1/2},~~
S_c = \frac{\pi v_c^2}{4}.
\eeq
The value of $\alpha$, $\beta$, $\gamma$, and $\delta$ can be found in Ref. \cite{Wei:2015ana}. Therefore the reduced temperature $\tilde T$ and Gibbs free energy $\tilde G$ are
\beq
\tilde{T} &=&\frac{\beta^4 \tilde{S}^2 (2 \pi  \alpha \beta^2 \tilde{P} \tilde{S}+1) (2 \pi \alpha \beta^2 \tilde{P} \tilde{S}+3)-192}{2 \pi  \gamma \beta^3 \tilde{S}^{3/2} \sqrt{2 \pi \alpha \beta^2 \tilde{P} \tilde{S}+3} \sqrt{\beta^4 \tilde{S}^2 (2 \pi \alpha \beta^2 \tilde{P} \tilde{S}+3)+192}}\,,\\
\tilde{G} &=&\frac{ \beta ^2 \tilde{S} (9 \beta ^2 \tilde{S}-4 \pi  \alpha  \tilde{P} (\pi  \alpha  \beta ^6 \tilde{P} \tilde{S}^3-192))+1728}{24 \delta  \sqrt{\beta ^2 \tilde{S}} \sqrt{2 \pi  \alpha  \beta ^2 \tilde{P} \tilde{S}+3} \sqrt{\beta ^4 \tilde{S}^2 (2 \pi  \alpha  \beta ^2 \tilde{P} \tilde{S}+3)+192}}\,.
\eeq
Similar to the charged AdS black hole case, the $\tilde T$ and $\tilde G$ of the Kerr-AdS black hole are independent of $J$. We show the $\tilde{T}-\tilde{S}$ and $\tilde{G}-\tilde{T}$ diagrams in Fig. \ref{KerrTSGT}. The results are same as that of the RN-AdS black hole: the phase transition takes place for $P<P_c$ and disappears for $P>P_c$. The reduced coexistence curve is also plotted in Fig. \ref{KerrPT}.

\section{Critical behaviors of KN-AdS black holes}\label{Kerr-Newman}

In the last section, we clearly show that the critical behaviors of the charged RN-AdS black hole and rotating Kerr-AdS black hole are independent of the charge $Q$ angular momentum $J$, respectively. Here we would like to ask how does the KN-AdS black hole of nonvanishing $Q$ and $J$ behave?

\subsection{Thermodynamic quantities of KN-AdS black holes}
The metric of the KN-AdS black hole is
\beq\label{KNmetric}
ds^2 &=& - {\Delta_r \over \rho^2}\Bigr(\,\m{d}t \; - \; {a \m{sin} \over \Xi}^2\theta \,\m{d}\phi\Bigl)^2\;+\;{\rho^2 \over \Delta_r}\m{d}r^2\;+\;{\rho^2 \over \Delta_{\theta}}\m{d}\theta^2  \nonumber \\
&~&+\;{\m{sin}^2\theta \,\Delta_{\theta} \over \rho^2}\Bigr(a\,\m{d}t \; - \;{r^2\,+\,a^2 \over \Xi}\,\m{d}\phi \Bigl)^2,
\eeq
where
\beq
\rho^2& = & r^2\;+\;a^2\m{\cos}^2\theta, \nonumber\\
\Delta_r & = & (r^2+a^2)\Big(1 + {r^2\over l^2}\Big) - 2mr+q^2,\nonumber\\
\Delta_{\theta}& = & 1 - {a^2\over l^2} \, \m{\cos}^2\theta, \nonumber\\
\Xi & = & 1 - {a^2\over l^2}.
\eeq
The $U(1)$ potential reads
\be
A=-\frac{qr}{\rho^2}\left(\m{d}t-\frac{a\sin^2\!\theta}{\Xi}\m{d}\varphi\right)\,.
\ee
The thermodynamic quantities of the KN-AdS black hole are \cite{Caldarelli:1999xj}
\beq
S &=& \frac{\pi(r_+^2+a^2)}{\Xi}\,,\nonumber\\
T &=&\frac{r_+\left(1+\frac{a^2}{l^2}+3\frac{r_+^2}{l^2}-\frac{a^2+q^2}{r_+^2}\right)}{4\pi (r_+^2+a^2)}\,,\nonumber\\
V &=& \frac{2\pi}{3}\frac{(r_+^2+a^2)(2r_+^2l^2+a^2l^2-r_+^2a^2)+l^2q^2a^2}{l^2\Xi^2 r_+}\,,\nonumber\\
\Phi &=& \frac{qr_+}{r_+^2+a^2}\,,\nonumber\\
\Omega_H &=& \frac{a\Xi}{r_+^2+a^2}\,,\nonumber\\
\eeq
where $r_+$ denotes the radius of the black hole event horizon. Regarding the cosmological constant as a pressure $P$, one will get
\be
m =\frac{64 \pi ^2 a^4 P^2 q^2-48 \pi  a^2 P q^2+24 \pi  a^2 P r_+^2+9 a^2+24 \pi  P r_+^4+9 q^2+9 r_+^2}{18 r_+}.
\ee
The physical mass $M$, charge $Q$, and angular momentum
$J$ can be expressed by the parameters $m$, $q$, and $a$ as:
\be
M=\frac{m}{\Xi^2}\,,\quad Q=\frac{q}{\Xi}\,,\quad J=\frac{am}{\Xi^2}\,.
\ee
After a simple calculation, the temperature $T$, mass $M$, and thermodynamic volume $V$ can be expressed in terms of the entropy $S$, angular momentum $J$, and pressure $P$:
\beq
T &=& \frac{-12 \pi ^2 J^2+S^2 \Bigl(64 P^2 S^2+32 P S+3\Bigr)+16 \pi  P Q^2 S^2-3 \pi ^2 Q^4}{4 \sqrt{\pi } S^{3/2} \sqrt{12 \pi ^2 J^2 (8 P S+3)+\Sigma^2}}\,,\\
M &=& \frac{m}{\Xi^2}=\frac{\sqrt{12 \pi ^2 J^2 (8 P S+3)+\Sigma^2}}{6 \sqrt{\pi }\sqrt{S}}\,,\\
V &=& 4\sqrt{S}\frac{2592 \pi ^6 J^6 (8 P S+3) \Bigl(\pi  Q^2 (8 P S+3)+3 S\Bigr)}{3 \sqrt{\pi} \Sigma \Bigl(36 \pi ^2 J^2+\Sigma^2\Bigr)^2 \sqrt{12 \pi ^2 J^2 (8 P S+3)+\Sigma^2}} \nonumber\\
 &~& +4\sqrt{S}\frac{432 \pi ^4 J^4 \Bigl(\pi  Q^2 (8 P S+3)+2 S (4 P S+3)\Bigr) \Sigma^2}{3 \sqrt{\pi } \Sigma \Bigl(36 \pi ^2 J^2+\Sigma^2\Bigr)^2 \sqrt{12 \pi ^2 J^2 (8 P S+3)+\Sigma^2}} \nonumber\\
 &~& +4\sqrt{S}\frac{6 \pi ^2 J^2 \Sigma^4 \Bigl(S (8 P S+15)+3 \pi  Q^2\Bigr)+S \Sigma^6}{3 \sqrt{\pi } \Sigma \Bigl(36 \pi ^2 J^2+\Sigma^2\Bigr)^2 \sqrt{12 \pi ^2 J^2 (8 P S+3)+\Sigma^2}}\,,
\eeq
where
\be
\Sigma=S (8 P S+3)+3 \pi  Q^2.
\ee
Thus the Gibbs free energy $G$ reads
\be
G =M-T S =\frac{12 \pi ^2 J^2 (16 P S+9)-64 P^2 S^4+12 \pi  Q^2 S (4 P S+3)+27 \pi ^2 Q^4+9 S^2}{12 \sqrt{\pi } \sqrt{S} \sqrt{12 \pi ^2 J^2 (8 P S+3)+(S (8 P S+3)+3 \pi  Q^2)^2}}.
\ee
It will reduce to the RN-AdS and Kerr-AdS black hole cases with $J=0$ and $Q=0$, respectively.

\subsection{Critical points for Kerr-Newman-AdS black holes}

\begin{figure}
    \centering
    \subfigure[]{\label{alpha}
    \includegraphics[height=0.3\textwidth]{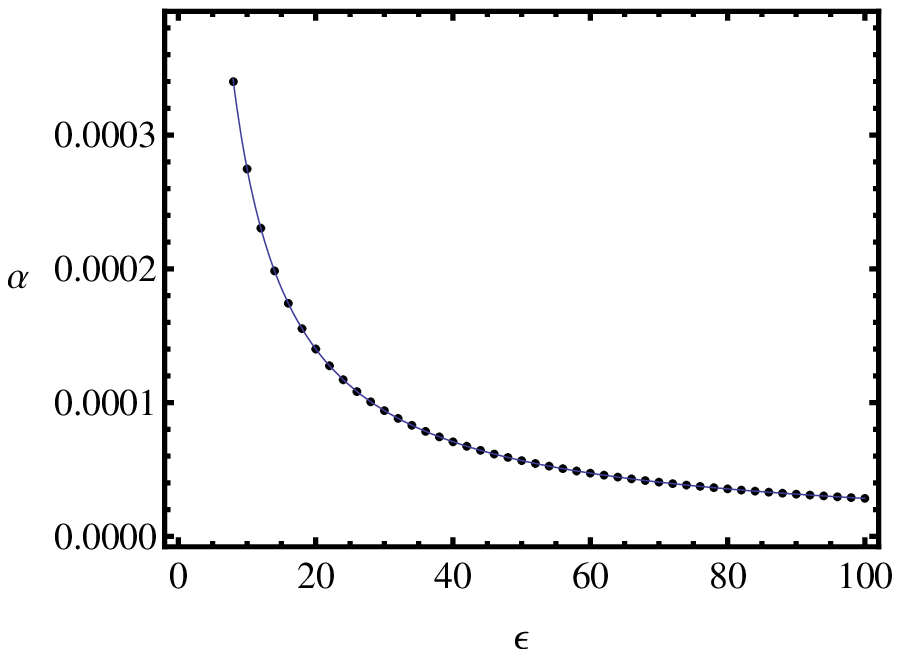}}
    \subfigure[]{\label{beta}
    \includegraphics[height=0.3\textwidth]{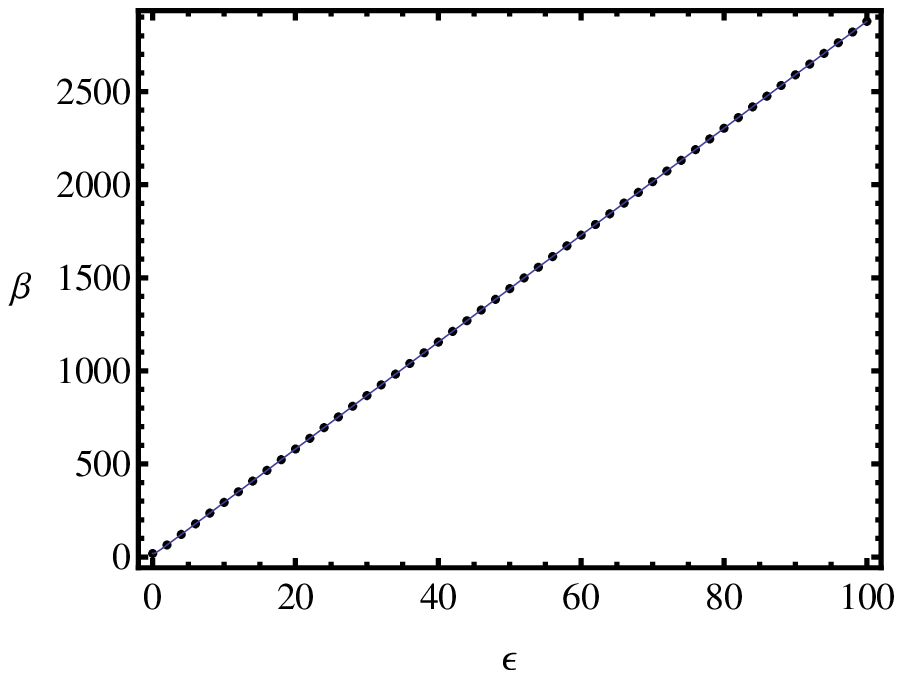}}
    \subfigure[]{\label{gamma}
    \includegraphics[height=0.3\textwidth]{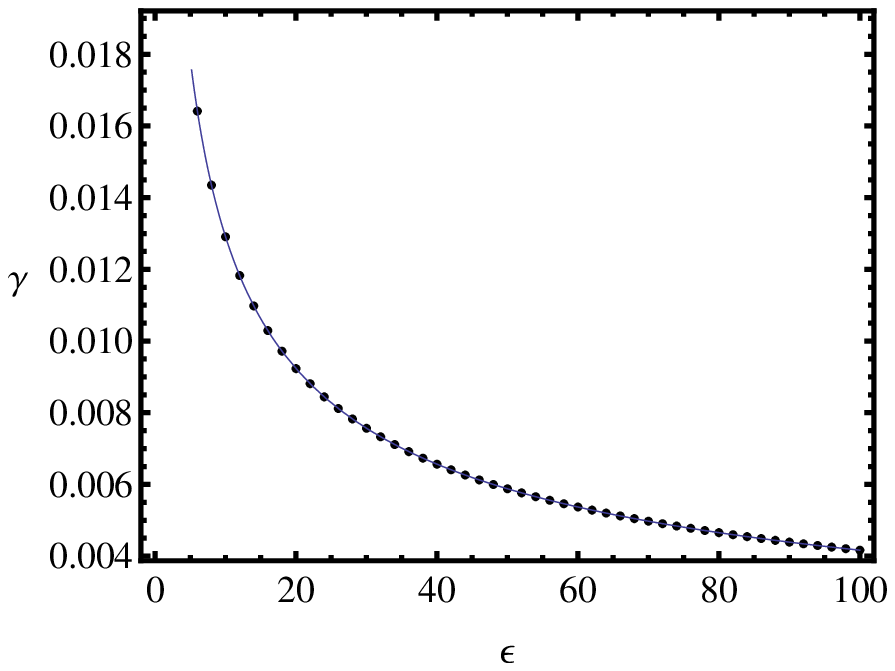}}
    \subfigure[]{\label{delta}
    \includegraphics[height=0.3\textwidth]{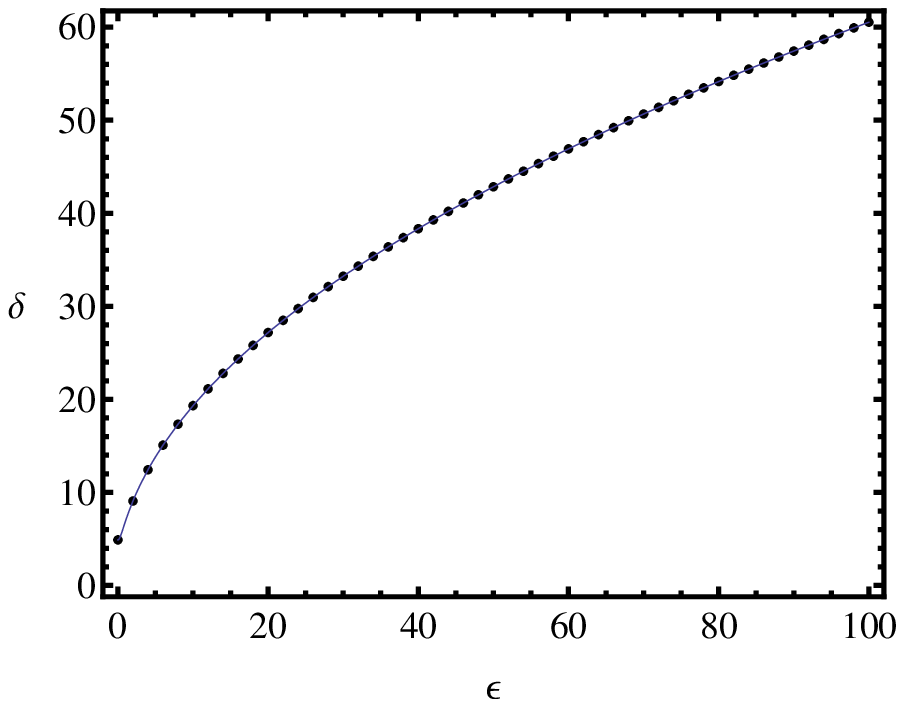}}
    \subfigure[]{\label{eta}
    \includegraphics[height=0.3\textwidth]{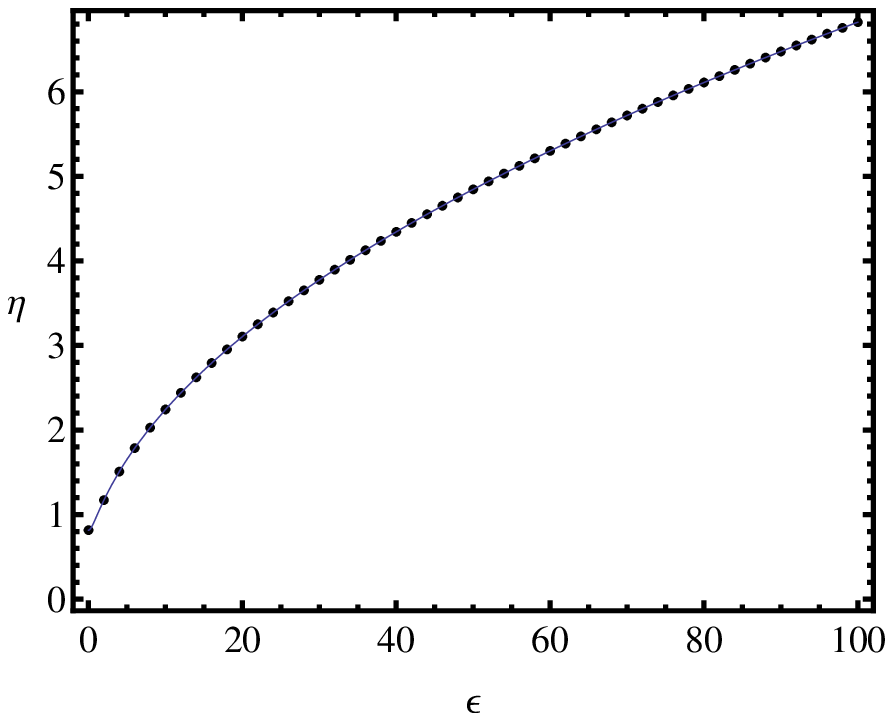}}
    \caption{Behaviors of the coefficients $\alpha$, $\beta$, $\gamma$, $\delta$, and $\eta$ as a functions of $\epsilon$. The dots denote the exact numerical results, and the solid lines are the fitting results with the forms shown in Eqs. (\ref{a})-(\ref{e}). We can see that they are highly consistent with each other. (a) $\alpha(\epsilon)$, (b) $\beta(\epsilon)$, (c) $\gamma(\epsilon)$, (d) $\delta(\epsilon)$, and (e) $\eta(\epsilon)$.}\label{Error}
\end{figure}

The critical point can be determined by the conditions $\partial_{S}T=\partial_{S,S}T=0$, which reduce to
\beq
&~& 144 \pi ^4 J^4 (32 P S+9)+24 \pi ^2 J^2 \left(3 \pi ^2 Q^4 (16 P S+9)+36 \pi  Q^2 S (4 P S+1)+S^2 (8 P S+3)^2 (16 P S+3)\right)\nonumber\\
&~& +\left(S (8 P S-1)+3 \pi  Q^2\right) \left(S (8 P S+3)+3 \pi  Q^2\right)^3=0\,,\label{pt1}\\
&~& 5184 \pi ^6 J^6 \left(512 P^2 S^2+288 P S+45\right)+144 \pi ^4 J^4 \left(9 \pi ^2 Q^4 \left(512 P^2 S^2+576 P S+135\right)\right.\nonumber\\
&~& \left.+72 \pi  Q^2 S \left(320 P^2 S^2+174 P S+27\right)+S^2 \left(-32768 P^4 S^4+20160 P^2 S^2+8640 P S+1053\right)\right)\nonumber\\
&~& +12 \pi ^2 J^2 \left(432 \pi ^3 Q^6 S \left(160 P^2 S^2+126 P S+27\right)+162 \pi ^2 Q^4 S^2 \left(960 P^2 S^2+560 P S+87\right)\right.\nonumber\\
&~& \left.-24 \pi  Q^2 S^3 \left(4096 P^4 S^4-9984 P^3 S^3-11520 P^2 S^2-3780 P S-405\right)+243 \pi ^4 Q^8 (32 P S+15)\right.\nonumber\\
&~& \left.+5 S^4 (8 P S+3)^4 (32 P S+9)\right)+\left(S (8 P S-3)+15 \pi  Q^2\right) \left(S (8 P S+3)+3 \pi  Q^2\right)^5=0 \label{pt2}.
\eeq
The above two equations will reduce to that of the RN-AdS or Kerr-AdS black hole when setting $J=0$ or $Q=0$. By solving it, one will get the analytic critical point. In Ref. \cite{Wei:2015ana}, we divided the thermodynamic quantities of a system into two classes, the universal and characteristic parameter. Since the $d$-dimensional singly spinning Kerr-AdS and charged RN-AdS black hole systems are single characteristic parameter thermodynamic systems, there exists analytic or exact critical point. Here we would like to apply the same technique to the KN-AdS black hole. However, for $J\neq0$ and $Q\neq0$, analytic or exact result could not be obtained, instead there is only the numerical result. After a simple analysis, we find that the KN-AdS black hole is a thermodynamic system with two characteristic parameters ($J$ and $Q$), thus its critical point must be in the following form
\beq
P_c = P_{c}(J, Q), \quad
S_c = S_{c}(J, Q),\quad
T_c = T_{c}(J, Q),\quad
v_c = v_{c}(J, Q),\quad
G_c = G_{c}(J, Q),\label{criticalpoint0}
\eeq
which means that the critical point are functions of $J$ and $Q$. Here we would like to define a new dimensionless AMC ratio $\epsilon$
\be
\epsilon=\frac{J}{Q^2}.
\ee
Thus for the KN-AdS black hole, we can convert the characteristic parameters $(J, Q)$ to $(\epsilon, J)$ or $(\epsilon, Q)$. Here we adopt the the latter one. Then using the dimensional analysis, the critical point can be expressed as
\beq
P_c = \frac{\alpha(\epsilon)}{Q^2}\,,\quad
S_c = \beta(\epsilon) \cdot Q^2\,,\quad
T_c = \frac{\gamma(\epsilon)}{Q}\,,\quad
v_c = \delta(\epsilon) \cdot Q\,,\quad
G_c = \eta(\epsilon) \cdot Q\,.\label{criticalpoint}
\eeq
The dimensionless coefficients $\alpha(\epsilon)$, $\beta(\epsilon)$, $\gamma(\epsilon)$, $\delta(\epsilon)$, and $\eta(\epsilon)$ are only the functions of the AMC ratio, and can be obtained by numerically solving Eqs. (\ref{pt1}) and (\ref{pt2}). It is worthwhile to point out that we can construct fitting forms for these functions. The construction must satisfy: i) when $\epsilon=0$, Eq. (\ref{criticalpoint}) reproduces the critical point of the RN-AdS black hole, and ii) when $\epsilon=\infty$, it gives the case of the Kerr-AdS black hole. Finally, we get the fitting forms for these functions
\beq
 \alpha&=& \frac{2.196 \epsilon ^5+2.728 \epsilon ^4+2.013 \epsilon ^3+0.9592 \epsilon ^2+0.2953 \epsilon +0.05070}{768.6 \epsilon ^6+1249 \epsilon ^5+1154 \epsilon ^4+712.9 \epsilon ^3+307.6 \epsilon ^2+89.05 \epsilon +15.29}\,,\label{a}\\
 \beta&=& \frac{739.4 \epsilon ^6+1089 \epsilon ^5+967.4 \epsilon ^4+582.3 \epsilon ^3+248.4 \epsilon ^2+71.82 \epsilon +12.86}{25.74 \epsilon ^5+33.00 \epsilon ^4+24.80 \epsilon ^3+12.09 \epsilon ^2+3.810 \epsilon +0.6820}\,,\label{b}\\
 \gamma&=& \frac{2.544 \epsilon ^{3/2}+7.722 \epsilon ^{5/2}+14.78 \epsilon ^{7/2}-0.9838 \epsilon ^3+3.548 \epsilon ^2+2.030 \epsilon +0.3348 \sqrt{\epsilon }+0.7765}{54.61 \epsilon ^{3/2}+78.99 \epsilon ^{5/2}-23.57 \epsilon ^{7/2}+354.1 \epsilon ^4+263.0 \epsilon ^3+117.8 \epsilon ^2+47.25 \epsilon +7.709 \sqrt{\epsilon }+17.93}\,,\label{c}\\
 \delta&=& \frac{155.0 \epsilon ^{7/2}-13.53 \epsilon ^3+81.17 \epsilon ^{5/2}+32.03 \epsilon ^2+31.47 \epsilon ^{3/2}+19.04 \epsilon +3.584 \sqrt{\epsilon }+9.854}{25.63 \epsilon ^3-2.241 \epsilon ^{5/2}+11.00 \epsilon ^2+5.219 \epsilon ^{3/2}+4.024 \epsilon +0.7236 \sqrt{\epsilon }+2.012}\,,\label{d}\\
 \eta&=& \frac{-64 \alpha ^2 \beta ^4+3 \pi ^2 \left(4 \epsilon ^2 (16 \alpha \beta +9)+9\right)+12 \pi  \beta  (4 \alpha \beta +3)+9 \beta ^2}{12 \sqrt{\pi } \sqrt{\beta } \sqrt{12 \pi ^2 \epsilon ^2 (8 \alpha \beta +3)+(\beta  (8 \alpha \beta +3)+3 \pi )^2}}\,.\label{e}
\eeq
The behaviors of $\alpha(\epsilon)$, $\beta(\epsilon)$, $\gamma(\epsilon)$, $\delta(\epsilon)$, and $\eta(\epsilon)$ are shown in Fig. \ref{Error}. The dots denote the exact numerical result, and the solid lines are obtained from the fitting forms (\ref{a})-(\ref{e}). From the figure, we can clearly see that our fitting forms are highly consistent with the numerical result. A detailed calculation shows that the relative deviation of these coefficients are less than $0.0001\%$. Moreover, we can see that all of them are monotone functions of the AMC ratio. $\alpha$ and $\gamma$ decrease with $\epsilon$, while $\beta$, $\delta$, and $\eta$ increase with it. In Refs. \cite{Caldarelli:1999xj, Dolan:2012jh}, the authors got the line of second-order critical point in the $J-Q$ plane with $l=1$. Equivalently, setting $P_c=3/8\pi$ in $P_c=\alpha(\epsilon)/Q^2$ in Eq. (\ref{criticalpoint}), we can obtain the similar line.

\subsection{Phase diagrams and coexistence curves}

We will study the detailed information of phase transition in this subsection. In the reduced parameter space, the temperature and Gibbs free energy are given by
\beq
\tilde T &=& \frac{\beta^2 \tilde{S}^2 \left(64 \alpha^2 \beta^2 \tilde{P}^2 \tilde{S}^2+32 \alpha \beta \tilde{P} \tilde{S}+3\right)+16 \pi  \alpha \beta^2 \tilde{P} \tilde{S}^2-3 \pi ^2 \left(4 \epsilon ^2+1\right)}{4 \sqrt{\pi } \gamma \left(\beta \tilde{S}\right)^{3/2} \sqrt{12 \pi ^2 \epsilon ^2 (8 \alpha \beta \tilde{P} \tilde{S}+3)+(\beta \tilde{S} (8 \alpha \beta \tilde{P} \tilde{S}+3)+3 \pi )^2}}\,,\\
\tilde G &=& \frac{-64 \alpha^2 \beta^4 \tilde{P}^2 \tilde{S}^4+3 \pi ^2 \left(4 \epsilon ^2 (16 \alpha \beta \tilde{P} \tilde{S}+9)+9\right)+12 \pi  \beta \tilde{S} (4 \alpha \beta \tilde{P} \tilde{S}+3)+9 \beta^2 \tilde{S}^2}{12 \sqrt{\pi } \eta \sqrt{\beta \tilde{S}} \sqrt{12 \pi ^2 \epsilon ^2 (8 \alpha \beta \tilde{P} \tilde{S}+3)+(\beta \tilde{S} (8 \alpha \beta \tilde{P} \tilde{S}+3)+3 \pi )^2}}\,.
\eeq
Since $\alpha$, $\beta$, $\gamma$, $\delta$, and $\eta$ are just the function of $\epsilon$, the reduced temperature and Gibbs free energy are only dependent of the AMC ratio $\epsilon$ rather than $J$ and $Q$. Taking $\epsilon=0$ or $\infty$, one will get the reduce temperature and Gibbs free energy for the RN-AdS black hole or Kerr-AdS black hole, respectively.

\begin{figure}[!htb]
  \centering
  \subfigure[]{\label{KNTS1}
    \includegraphics[height=0.3\textwidth]{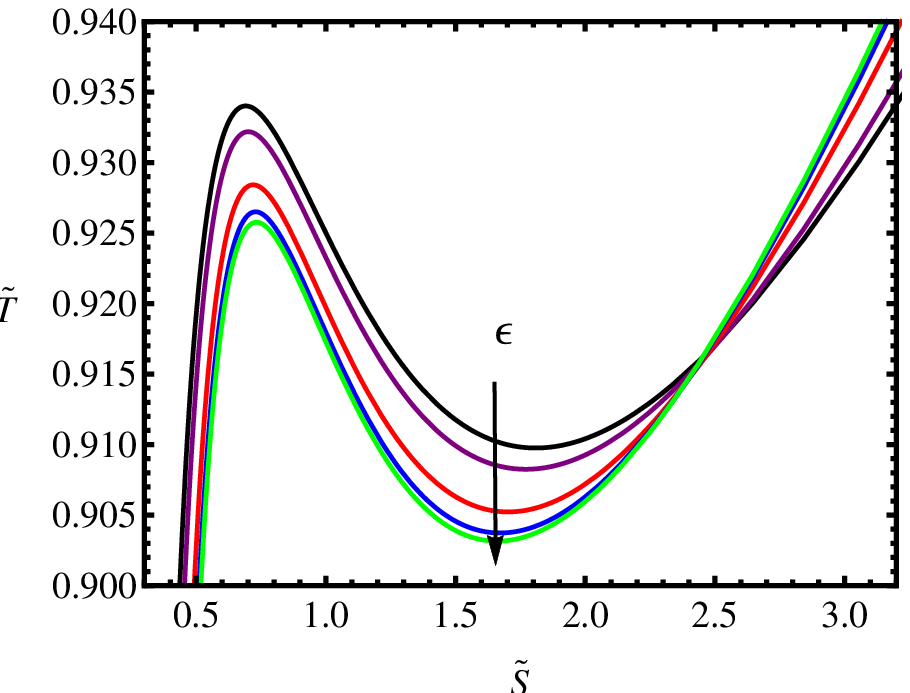}}
    \subfigure[]{\label{KNGT1}
    \includegraphics[height=0.3\textwidth]{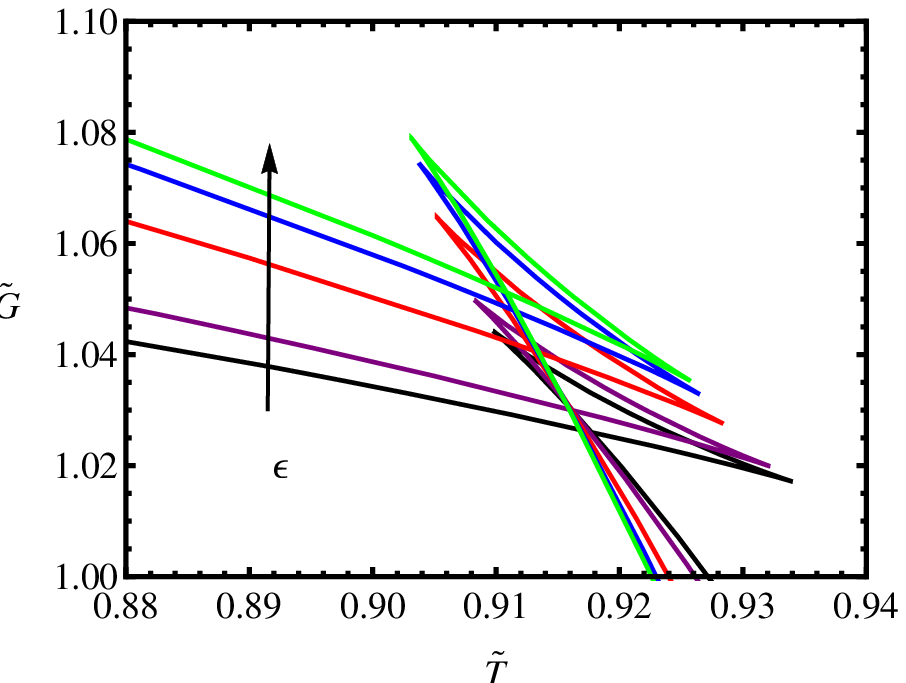}}
  \subfigure[]{\label{KNTS2}
    \includegraphics[height=0.3\textwidth]{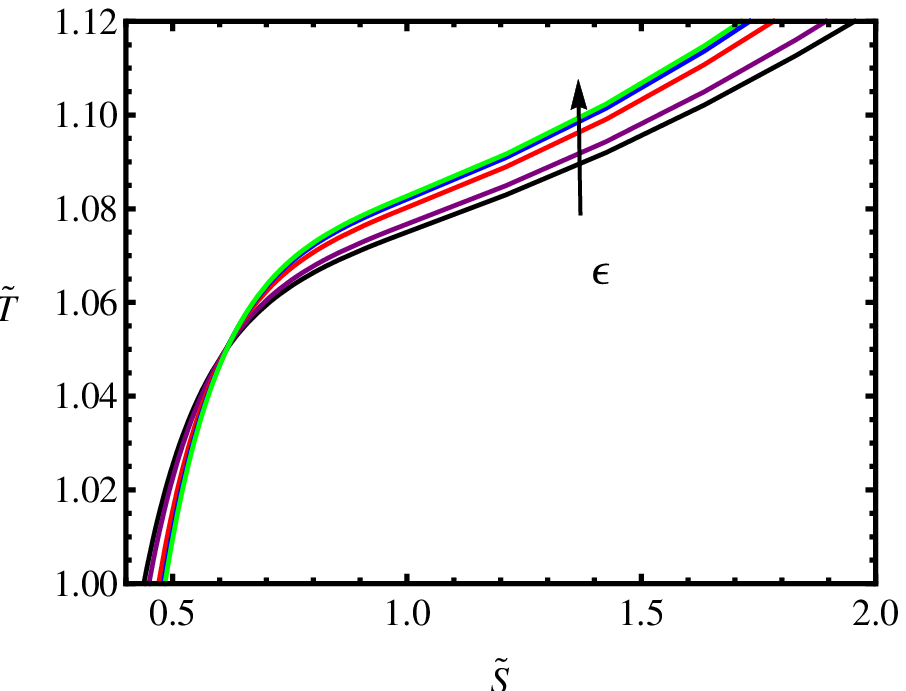}}
    \subfigure[]{\label{KNGT2}
    \includegraphics[height=0.3\textwidth]{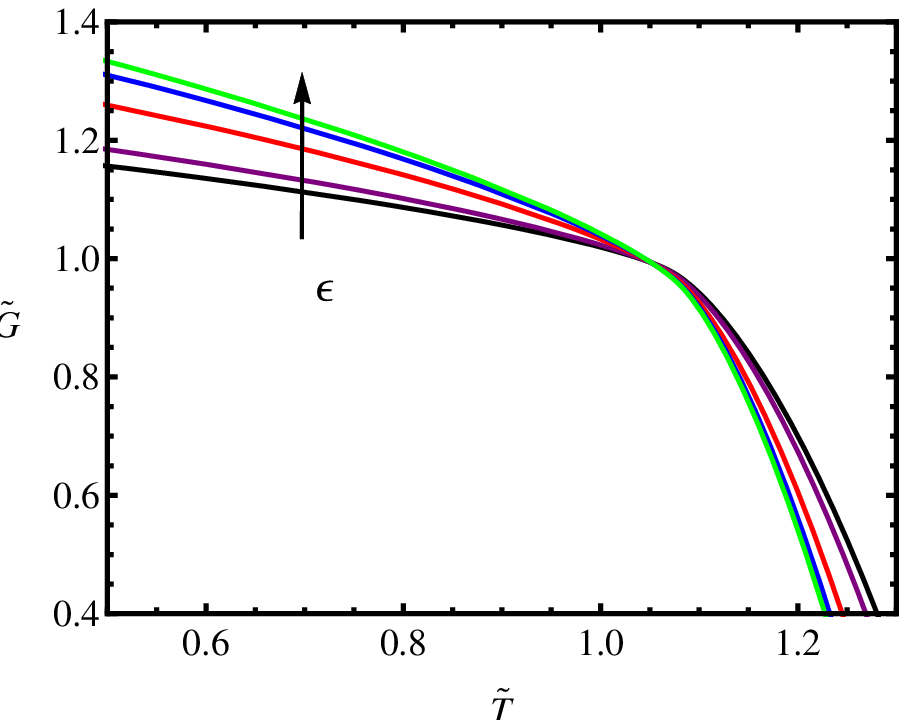}}
\caption{Reduced temperature and Gibbs free energy for the KN-AdS black holes with $\epsilon$=0, 1, 5, 10, and $\infty$, marked with black, purple, red, blue, green colors. Black arrows indicate increasing $\epsilon$. (a) $\tilde{T}$ vs. $\tilde{S}$ with $\tilde P$=0.8. (b) $\tilde{G}$ vs. $\tilde{T}$ with $\tilde P$=0.8. (c) $\tilde{T}$ vs. $\tilde{S}$ with $\tilde P$=1.2. (d) $\tilde{G}$ vs. $\tilde{T}$ with $\tilde P$=1.2. One could see that for $\tilde P<1$, the reduced temperature exhibits an oscillatory behavior, and Gibbs free energy exhibits a swallow tail behavior. While for $\tilde P>1$, these behaviors disappear.}\label{KNGT0}
\end{figure}

We plot $\tilde{T}$ and $\tilde{G}$ in Fig. \ref{KNGT0} with different AMC ratio $\epsilon$. For $\tilde{P}=0.8$, the reduced temperature displays the oscillatory behavior and the Gibbs free energy demonstrates the swallow tail behavior. With the increase of $\epsilon$, the local maximum and minimum points of $\tilde{T}$ decrease, and the intersection point of $\tilde{G}$ is also shifted toward lower $\tilde{T}$ and higher $\tilde{G}$. When $\tilde{P}=1.2$ the oscillatory behavior of the isobaric line and swallow tail behavior disappear. All the figures confirm a $\epsilon$-dependent behavior.

\begin{figure}[!htb]
    \subfigure[]{\label{KNPT1}
    \includegraphics[height=0.4\textwidth]{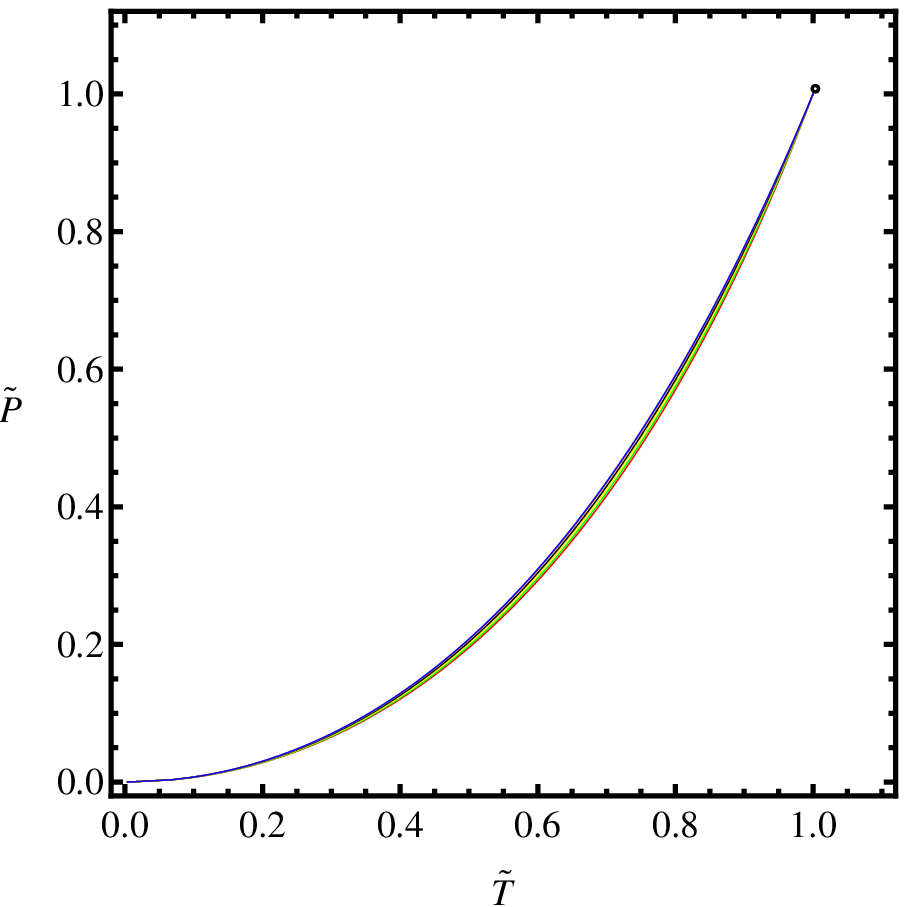}}
    \subfigure[]{\label{KNPT2}
    \includegraphics[height=0.4\textwidth]{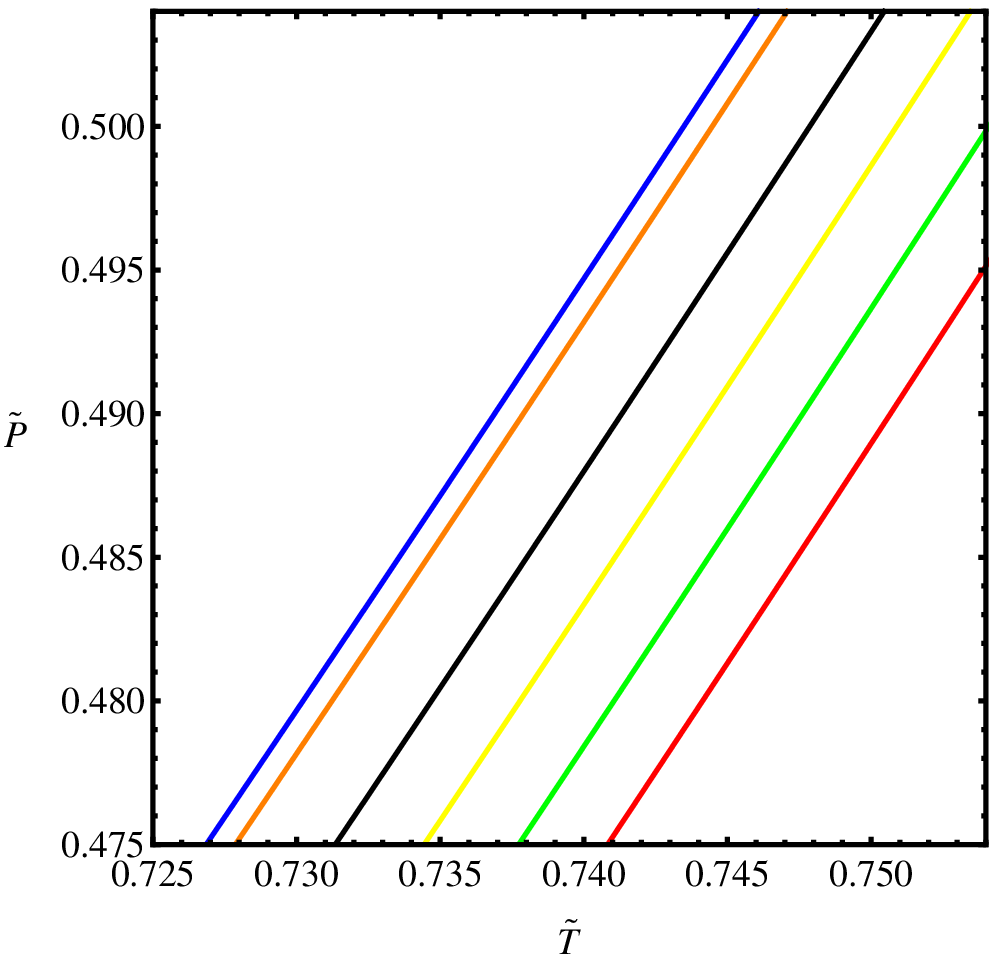}}
     \caption{(a) Coexistence curves in the reduced parameter space for the KN-AdS black hole with $\epsilon$=0, 0.5, 1, 2, 10, and $\infty$ from bottom to top. (b) Magnified coexistence curves of (a). The bottom red line and top blue line are coexistence curves for the RN-AdS black hole ($\epsilon=0$) and Kerr-AdS black hole ($\epsilon=\infty$), respectively.  When increasing $\epsilon$ from 0 to $\infty$, the coexistence curve of the KN-AdS black hole will continuously change from that of the RN-AdS to the Kerr-AdS black hole.}\label{KNPT}
\end{figure}

The oscillatory behavior of the reduced temperature and the swallow tail behavior of the reduced Gibbs free energy imply that there exists a first-order phase transition, which is known as the small-large black hole phase transition. The coexistence point, where the small and large black holes coexist at, can be obtained by constructing the equal area law on the isobaric line or determining the intersection point of the Gibbs free energy. Since $\tilde T$ and $\tilde G$ are only dependent of the AMC ratio $\epsilon$, the coexistence curve is dependent of $\epsilon$ rather than $J$ and $Q$. For each fixed $\epsilon$, the coexistence curve can be numerically solved. Applying the same fitting technique in Ref. \cite{Wei:2014qwa}, we can get the highly accurate fitting formula for the coexistence curve. The fitting form is as follows:
\be
\tilde P= \sum^{10}_{i=0}a_i \tilde T^i,~~ \tilde T \in (0,1)\,,
\ee
where $a_0-a_{10}$ are the fitting coefficients, and they are function of $\epsilon$. A general result gives $a_0=a_1=0$ for any $\epsilon$. For examples, we list the values of the coefficients $a_i$ ($2 \leq i\leq 10$) for different $\epsilon$ in Table \ref{tb1}. These results are very useful for further study on the thermodynamic property of the system varying along the coexistence curve. Moreover, we plot the fitting curve in Fig. \ref{KNPT}. For different values of $\epsilon$, there only exists a tiny difference. From bottom to top, the ratio $\epsilon=0, 0.5, 1, 2, 10, \infty$, respectively. So the bottom one is for the RN-AdS black hole case, and the top one for the Kerr-AdS black hole one. Interestingly, for the KN-AdS black hole with nonzero and finite $\epsilon$, its coexistence curve is located between that of the RN-AdS and Kerr-AdS ones. And with increasing $\epsilon$ from 0 to $\infty$, its coexistence curve will continuously change from that of the RN-AdS black hole to the Kerr-AdS black hole.

Moreover, we also numerically check the relative deviations of the coexistence curves between a small $\epsilon$ one and that of the RN-AdS black hole one. The result suggests that when $\epsilon\leq 0.01$, the relative deviation will no more larger than $0.1\%$. On the other hand, the relative deviation will fall in the same precision between a KN-AdS black hole with $\epsilon\geq100$ and a Kerr-AdS black hole. Therefore, one can believe that it is small enough for $\epsilon=0.01$ to describe a RN-AdS black hole case and large enough for $\epsilon=100$ to describe a Kerr-AdS black hole case.

\begin{table}[!htb]
    \tabcolsep 0pt
    \caption{Values of the coefficients $a_i$ in the fitting formula of the coexistence curve for different values of $\epsilon$.}
    \vspace*{-12pt}
    \begin{center}
    \def\temptablewidth{0.95\textwidth}
    {\rule{\temptablewidth}{1pt}}
    \begin{tabular*}{\temptablewidth}{@{\extracolsep{\fill}}cccccccccc}
    $\epsilon$ & $a_2$ & $a_3$ & $a_4$ & $a_5$ & $a_6$ & $a_7$ & $a_8$ & $a_9$ & $a_{10}$\\   \hline
    0  & 0.669854& 0.143639& 0.310866& -0.850256& 2.09597& -3.08471& 2.85851& -1.49106& 0.347242 \\
    0.01& 0.670168& 0.141295& 0.322274& -0.884237& 2.16061& -3.16342& 2.91788& -1.51634& 0.351883 \\
    0.5  & 0.678829 & 0.168124& 0.166430& -0.324466& 0.864016& -1.26953& 1.19729& -0.631584& 0.150939  \\
    1  &0.689774& 0.181979& 0.0883886& -0.0482319& 0.212793& -0.313865& 0.320250& -0.176552& 0.0454712  \\
    2  &0.700909& 0.189414& 0.0498247& 0.0787816& -0.0874291& 0.116773& -0.0705630& 0.0231946& -0.000914686  \\
    10  &0.714042& 0.195529& 0.0203609& 0.165749& -0.291314& 0.394774& -0.312537& 0.140058& -0.0266872 \\
    100  &0.717508& 0.197592& 0.0105573&  0.193513 & -0.352448 &  0.473886&  -0.377497 & 0.169463 & -0.0326054 \\
    $\infty$  &0.717984& 0.197176& 0.0125476& 0.187503& -0.342135& 0.462037& -0.369272& 0.166198& -0.0320660
    \end{tabular*}
    {\rule{\temptablewidth}{1pt}}
    \end{center}
    \label{tb1}
\end{table}

In a word, in the reduced parameter space, the temperature, Gibbs free energy, and coexistence curve for the KN-AdS black hole are only dependent of the AMC ratio $\epsilon$ rather than the angular momentum $J$ and charge $Q$.

\section{conclusions}\label{conclusions}

In this paper, we have studied the critical phenomena for the KN-AdS black holes in the extended phase space. We firstly reviewed the critical behaviors for the RN-AdS black hole and Kerr-AdS black hole. In the reduced parameter space, one can easily find that the critical behavior is charge independent for the RN-AdS black hole and angular momentum independent for Kerr-AdS black hole. The reason is that these black hole systems are single characteristic parameter thermodynamic systems.

For a KN-AdS black hole with nonvanishing $Q$ and $J$, it is a two-characteristic-parameter thermodynamic system. Therefore, according to the dimensional analysis, its critical point depends both on the parameters $Q$ and $J$. And it seems at first sight that the state equation and the coexistence curve are also dependent of $Q$ and $J$. Before further analysis, we defined a dimensionless AMC ratio $\epsilon$. Then the critical point can be put in the forms of Eq. (\ref{criticalpoint}). Using these forms, we obtained the temperature and Gibbs free energy in the reduced parameter space, and found that they are only dependent of the ratio $\epsilon$, which implies that the state function and coexistence curve are also dependent of $\epsilon$ rather than $Q$ and $J$.

We numerically solved the coefficients of the critical point. And highly accurate fitting formula for them are also obtained. The relative deviation between the exact numerical results and fitting results are less than $0.0001\%$. Moreover, we also gave the fitting formula for the coexistence curve with several $\epsilon$ shown in Table \ref{tb1}. Its behavior was also plotted in Fig. \ref{KNPT}. For $\epsilon=0$, and $\infty$, they are just that for the RN-AdS black hole and Kerr-AdS black hole. And varying $\epsilon$ from 0 to $\infty$, the coexistence curve for the KN-AdS black hole continuously change from the RN-AdS case to the Kerr-AdS case.

Before ending this paper, we would like to make a few comments. In Ref. \cite{Wei:2015ana}, we examined the property for the thermodynamic quantities of a black hole system. And then we divided these quantities into two classes, i.e., the universal parameter and characteristic parameter. For the single characteristic parameter thermodynamic system, we argued that the form of the critical points can be uniquely determined by dimensional analysis. In the reduced parameter space, the coexistence curve is independent of its characteristic parameter. Under this viewpoint, we successfully studied the critical behavior for the $d$-dimensional singly spinning Kerr-AdS black hole system, which is a typical single characteristic parameter system. However this viewpoint becomes invalid for the two or more characteristic parameters system. The KN-AdS black hole system is a two-characteristic-parameter system. We in this paper showed that its critical point depend on the two characteristic parameters $Q$ and $J$. However, in the reduced parameter space, the state equation, Gibbs free energy, and coexistence curve are dependent of only one parameter, i.e., the ratio $\epsilon$, rather than $Q$ and $J$. These results may guide us to study the critical phenomena of two-characteristic-parameter thermodynamic systems in future work.

\section*{Acknowledgements}
This work was supported by the National Natural Science Foundation of China (Grants No. 11205074 and No. 11375075).

\end{document}